\documentclass[reprint,amsmath,amssymb,aps, prc]{revtex4-2}

\usepackage{graphicx}
\usepackage{bm}
\usepackage[colorlinks=true,citecolor=blue,linkcolor=blue,urlcolor=blue]{hyperref}
\begin{document}

\title{Constrained functional priors for Bayesian inference of hot QCD matter}
\author{G. Guimarães}
\email{gabriel.alg@usp.br}

\author{L. Perin}
\email{luizaperin@usp.br}

\author{M. Luzum}
\email{mluzum@usp.br}
\affiliation{Instituto de Física, Universidade de São Paulo, R. do Matão, 1371, Brazil, 05508-090}

\begin{abstract}

Bayesian analyses of heavy-ion collisions rely on functional inputs, such as temperature-dependent transport coefficients and the equation of state, whose prior specification remains a significant source of uncertainty. Standard approaches employ low-dimensional parametrizations that can impose artificial correlations and leave the resulting inference sensitive to the assumed functional form.

We develop a nonparametric framework for constructing priors over such functions using Gaussian processes, tailored to emulator-based inference in heavy-ion phenomenology. Truncated Karhunen–Loève expansions yield optimized finite-dimensional representations suitable for use as emulator inputs. We investigate how physical constraints can be incorporated into these priors, demonstrating that rejection-based methods produce non-Gaussian measures and induce nontrivial statistical dependences in the expansion coefficients. By contrast, pushforward constructions enforce the constraints while preserving a tractable Gaussian measure in a latent space.

These results provide a systematic framework for incorporating physical constraints into functional priors and clarify their effects on the statistical structure of the finite-dimensional representations used in Bayesian inference.

\end{abstract}

\maketitle

\section{Introduction}
Heavy-Ion Collisions (HIC) have been one of the main sources of experimental guidance for the study of the strong interaction under extreme conditions \cite{Busza_2018, arslandok2023hotqcdwhitepaper}. These collisions produce an extremely hot but short-lived medium, and determining its properties has been an effort of great interest for both the theoretical and experimental communities. Early studies established the qualitative picture of heavy-ion collisions, including the formation of the quark–gluon plasma (QGP) and its strongly coupled, fluid-like behavior \cite{kolb2003hydrodynamicdescriptionultrarelativisticheavyion}. With theoretical and experimental advances, our understanding of these systems has become increasingly quantitative and precise. In particular, comparisons of hydrodynamic simulations with collider data demonstrated the necessity of non-trivial transport coefficients, and led to increasingly precise estimates of those quantities \cite{Romatschke_2007,Song_2008,Luzum:2008cw,Niemi_2011,Luzum_2013, Bernhard_2015,Bernhard_2016,Everett_2021,mankolli2026longitudinaldynamicslargesmall}.

A crucial step towards this ``precision era'' of HIC was the adoption of Bayesian inference methods in the field \cite{Novak_2014, Sangaline_2016, Bernhard_2015, Bernhard_2016}. The Bayesian approach is advantageous because it provides a principled way to account for not only the experimental sources of uncertainty but also the theoretical ones, which cannot be neglected in this case. The modeling of these collisions is a great challenge, and there is no first-principles approach that is successful for realistic collisions. The most successful descriptions are numerical simulations in which there is a succession of several distinct stages, each with an effective description suited to the current conditions of the system. The most ubiquitous feature of these models is, as suggested earlier, the presence of a hydrodynamic stage.

The community has developed a more or less standard framework for Bayesian inference to deal with domain-specific challenges, especially the high computational cost of numerical simulations \cite{Paquet_2024}. Its central ingredients are the use of Markov Chain Monte Carlo (MCMC) methods to sample the posterior distribution over model parameters, and the use of Gaussian-process emulators to evaluate the likelihood without running the full simulation at every MCMC step. This has motivated several recent developments in surrogate modeling for heavy-ion phenomenology, including multi-fidelity Gaussian-process emulators \cite{ji2024graphicalmultifidelitygaussianprocess, Liyanage_2022} and additive multi-index Gaussian processes for multi-physics QGP simulations \cite{li2023additivemultiindexgaussianprocess}. Related structured Gaussian-process models have also been developed more broadly for expensive computer-code emulation \cite{tang2023hierarchicalshrinkagegaussianprocesses}. However, while many MCMC algorithms used in practice are formulated for finite-dimensional parameter spaces, several quantities of interest in heavy-ion collisions are naturally functions, such as the temperature-dependent transport coefficients $\eta/s(T)$ and $\zeta/s(T)$. This motivates inference strategies that treat such objects as functions, rather than only through a small number of predetermined parameters.

This mismatch can be resolved in two ways: by using MCMC algorithms that are well-posed in the infinite-dimensional limit or by restricting to a finite-dimensional subspace for the inference. For practical reasons that will be discussed later, the former is not yet feasible in the context of realistic simulations, and the latter is the usual approach. Finite-dimensional parametrizations have been widely employed for inference, but they suffer from some well-known limitations. The restricted functional forms are inherently rigid and can induce spurious long-range correlations in the inferred functions \cite{PhysRevC.108.L011901, Paquet_2024, Legred_2022, mankolli2026longitudinaldynamicslargesmall}. Besides that, the resulting posteriors can depend strongly on the chosen parametrization, in the sense that apparent constraints may reflect the imposed functional structure rather than genuine information from the data. This phenomenon is commonly referred to as prior dominance.

In this paper, we propose a methodology, based on random process theory, to specify priors over infinite-dimensional function spaces and parametrize them in an optimal and mathematically motivated way that mitigates both issues mentioned in the previous paragraph. Following previous work suggesting the use of random processes as priors for inference over functions in HIC \cite{gong2024gaussianprocessgenerativemodel, PhysRevC.108.L011901, Paquet_2024}, we construct priors with different degrees of physical motivation for functions of interest and compute parametrizations that are optimal in the $L^2$ sense with respect to the prior measure. We quantify the effects of two strategies to impose physical constraints in the prior distribution, which we refer to as rejection sampling and pushforward constructions and discuss the benefits and limitations of each approach. The flexibility of our method provides a way to incorporate several kinds of physical information, with varying degrees of rigidity, in the prior and the construction of parametrizations for the resulting prior distributions that are readily applicable to state-of-the-art Bayesian analyses. This work focuses on the construction and statistical characterization of functional priors, rather than on a full end-to-end calibration to experimental data. Our goal is to develop and compare prior constructions and finite-dimensional representations that are compatible with existing emulator-based Bayesian workflows, and to clarify their practical implications for future heavy-ion collision analyses.

\section{Methodology}
A crucial step for Bayesian inference is specifying prior knowledge about the object of inference in the form of a probability distribution, the prior distribution. The standard approach to specify priors over functions in HIC phenomenology has been to choose a parametrization with a few free parameters and use simple probability distributions (usually uniform) for them. The prior knowledge is introduced mainly in two ways: the choice of parametrization, which may or may not reflect well-known features, and the choice of the probability distribution of the parameters. 

Another approach has recently gained attention in the field, following its successful application in other contexts, namely the use of random processes to specify these distributions \cite{PhysRevC.108.L011901, gong2024gaussianprocessgenerativemodel}. Continuous random processes are, in essence, probability distributions over spaces of functions and can therefore be used as priors. A useful prior should assign non-negligible probability mass to physically plausible functions, including the true one.

The use of random processes as priors for functions offers both conceptual and practical advantages. They are defined on infinite-dimensional spaces, which is natural for functions. Although in practice computer simulations always use finite-dimensional representations, these methods are still valuable because they can be designed to be robust under grid refinement. They also allow finer control of the correlations in the functions we want to infer and provide more flexibility than traditional parametrizations. 

GP priors have been successfully used for EoS inference with neutron-star observations \cite{Landry:2018prl, Essick:2019ldf, Landry:2020vaw, Mroczek:2023zxo}, for inference of Parton Distribution Functions \cite{Candido:2024hjt, medrano2026gaussianprocessesinferringparton}, for probing new physics in Cosmology \cite{Sabogal_2024}, for inferring wavefunction representations in quantum many-body physics \cite{PhysRevX.10.041026} and more. Recently, this strategy has been adopted by researchers in heavy-ion phenomenology. In \cite{gong2024gaussianprocessgenerativemodel}, a GP prior was constructed for the $\mu_B = 0$ case and the sensitivity of observables in state-of-the-art simulations to different samples of the EoS was demonstrated, while in \cite{PhysRevC.108.L011901} a GP prior was used for inference of the jet quenching function $\hat{q}(T)$.

We start the discussion with an introduction to the necessary theory of random processes.

\subsection{Random processes and the KL expansion}

As argued, random processes provide a natural framework for modeling uncertainty in functional quantities. Nearly every field of science has benefited from the application of models with stochastic elements, and this has motivated the development of a rich mathematical theory of random processes. In this work, we restrict our attention to processes with continuous sample paths, as their realizations must be continuous functions.

Consider a random process $X = \{ X(t)\}$, with $t \in [a, b] \subset \mathbb{R}$ which is second-order (meaning that it has finite variance), with zero mean and continuous covariance function $k(s,t)$. It can be decomposed as
\begin{equation}
    X(t) = \sum_{k=1}^\infty Z_k e_k(t) \label{eq1}
\end{equation}
with convergence in $L^2([a,b])$, where $Z_k$ are random variables and $e_k$ is a set of orthonormal functions that forms a basis of the relevant Hilbert space \cite{Rasmussen2006gaussian, ramsayFDA}. Furthermore, if we define the functions to be the eigenfunctions of the covariance integral operator $\hat{C}$ i.e. the solutions $e_k(t)$ of 
\begin{equation}
    [\hat{C} e_k](s) \equiv\int_a^b k(s,t) e_k(t) dt = \lambda_k e_k(s) \label{eq2}
\end{equation}
where $\lambda_k$ are the associated eigenvalues, then all the random variables $Z_k$ are uncorrelated (not necessarily independent), and it can be shown that truncations of the infinite sum in that basis yield finite linear parametrizations of the random process that minimize the average integrated squared error in the representation $\epsilon_N^2$:
\begin{equation}
    \epsilon_N^2 = \mathbb{E}\left\{ \int_a^b \left| X(t) -\sum_{k=1}^N Z_k e_k(t)  \right|^2 dt \right\}
\end{equation}
In other words, truncations of equation \eqref{eq1} are the optimal linear parametrizations in the $L^2$ sense for a given number of parameters $N$. We will refer to this expansion as the Karhunen-Loève (KL) expansion, after the famous Karhunen-Loève theorem that guarantees its existence. In a sense, this procedure is the infinite-dimensional analogue of Principal Component Analysis (PCA), since the KL modes defined by equation \eqref{eq2} form a basis that ``diagonalizes'' the covariance operator $\hat{C}$. For this reason, it is commonly called functional PCA. The KL expansion has been used for emulation in computer experiments with functional or high-dimensional outputs, where simulator predictions are projected onto a small number of dominant modes   \cite{ ji2024graphicalmultifidelitygaussianprocess}. Related KL-based representations have also been used to construct reduced-rank and structured Gaussian-process models \cite{tang2023hierarchicalshrinkagegaussianprocesses} and to parametrize random fields in Bayesian inverse problems \cite{Cotter_2013,URIBE2020112632}.

The random coefficients $Z_k$ can be decomposed in the following way
\begin{equation}
    Z_k =  \sqrt{\lambda_k} \xi_k  
\end{equation}
where $\xi_k$ is a random variable with unit variance and $\lambda_k$ is the variance contained in the mode $e_k$ (in the random process). The covariance function does not determine the joint distribution of $\xi_k$, and processes with the same covariance function can have distinct distributions due to higher correlations.

In this decomposition, the covariance function can be written in terms of the eigenfunctions and eigenvalues as
\begin{equation}
k(s,t) = \sum_{i} \lambda_i e_i(s) e_i(t). \label{eq5}
\end{equation}
In coefficient space,
\begin{equation}
\mathbb{E}[Z_k Z_l] = \lambda_k \delta_{kl}
\end{equation}
which shows that the modes are uncorrelated, while the full joint distribution of the $\xi_k$ is not fixed by the covariance alone. Although we restricted attention to scalar-input, scalar-output GPs, the Karhunen–Loève expansion extends to higher-dimensional inputs and to vector-valued outputs, with the KL modes defined over the input domain and, in the multi-output case, taking values in the output space.

An interesting case where the distribution of the random variables $\xi_k$ is known is for Gaussian processes, where the coefficients follow Gaussian distributions of zero mean and unit variance, which we denote by $\xi_k \sim \mathcal{N}(0,1)$ for each $k$, and the distributions are not only uncorrelated but completely independent. Special cases of covariance functions have analytical KL modes, but there are numerical routines that allow the calculation of modes in the general case. Analytical KL modes are known for some specific covariance kernels and choices of domain, measure, and boundary conditions. In general, however, the modes must be approximated numerically by solving the corresponding covariance-operator eigenvalue problem (see Appendix A).

Interestingly, similar constructions have appeared implicitly in the heavy-ion literature, where principal component analyses of fluctuating initial conditions are used to define uncorrelated modes of the initial state profiles \cite{Borghini:2022iym, Borghini:2024ekn, Krupczak:2025gwg}. While not always formulated in terms of random process theory, these approaches are closely related to the KL decomposition discussed here. The distributions of mode coefficients reported in these works are approximately Gaussian, with small deviations in the lowest modes, suggesting that the underlying ensembles of initial conditions are close to Gaussian random fields. A quantitative characterization of these deviations, and their physical origin, would be an interesting direction for future work. 

A notable difference, however, is that these analyses involve random fields defined over a two-dimensional transverse plane, rather than one-dimensional functions. For typical covariance kernels, the associated KL spectra decay more slowly as the dimensionality of the input space increases. As a result, a significantly larger number of modes is required to achieve a comparable level of accuracy. This is consistent with the relatively large number of principal components needed to reach the desired precision in such studies. A similar effect is expected to arise when building priors for the finite-density equations of state, which depend on multiple thermodynamic variables, where the increased dimensionality of the input space should likewise lead to a slower decay of the KL spectrum and a corresponding increase in the number of modes required for accurate representations.

This framework also provides a principled way to interpret studies in which PCA is applied to observables differential in transverse momentum \cite{Bhalerao_2015, Mazeliauskas_2015, Mazeliauskas_2016, Hippert_2020}, clarifying how the choice of binning relates to finite-dimensional approximations of the underlying Karhunen–Loève decomposition, as well as to the statistical properties of the ensembles being analyzed.

\subsection{Gaussian process regression}

Gaussian processes (GPs) are a class of random processes defined by the property that any finite collection of function values follows a multivariate Gaussian distribution \cite{Rasmussen2006gaussian}. For simplicity, we restrict our attention to GPs with input $\mathbf{x}$ from a subset of $\mathbb{R}^D$ and scalar output. A GP is fully specified by a mean function $m(\mathbf{x})$ and a covariance function $k(\mathbf{x},\mathbf{x}')$, defined as
\begin{align}
m(\mathbf{x}) &= \mathbb{E}[f(\mathbf{x})], \\
k(\mathbf{x}, \mathbf{x}') &= \mathbb{E}\big[(f(\mathbf{x}) - m(\mathbf{x}))(f(\mathbf{x}') - m(\mathbf{x}'))\big].
\end{align}
We denote this as
\begin{equation}
f(\mathbf{x}) \sim \mathcal{GP}(m(\mathbf{x}), k(\mathbf{x}, \mathbf{x}')).
\end{equation}

In Gaussian process regression, we infer a latent function from noisy observations
\begin{equation}
y_i = f(\mathbf{x}_i) + \epsilon_i,
\qquad
\boldsymbol{\epsilon} \sim \mathcal{N}(\mathbf{0}, \Sigma),
\end{equation}
where $\Sigma$ is the noise covariance matrix. The commonly used case of independent homoscedastic Gaussian noise corresponds to
\begin{equation}
\Sigma = \gamma^2 I,
\end{equation}
where $\gamma^2$ is the noise variance at each observation point.

Given training data $\{(\mathbf{x}_i, y_i)\}_{i=1}^n$, the predictive distribution at test inputs $X_*$ is Gaussian,
\begin{equation}
\mathbf{f}_* \sim \mathcal{N}(\bar{\mathbf{f}}_*, \operatorname{cov}(\mathbf{f}_*)),
\end{equation}
with mean and covariance
\begin{align}
\bar{\mathbf{f}}_* & =
m(X_*)  \nonumber \\ &\quad + K(X_*,X)
\bigl[K(X,X)+\Sigma\bigr]^{-1}
\bigl[\mathbf{y}-m(X)\bigr],\\
\operatorname{cov}(\mathbf{f}_*) &= K(X_*, X_*) \nonumber \\
&\quad - K(X_*, X)\bigl[K(X, X) + \Sigma\bigr]^{-1}K(X, X_*).
\end{align}

Here $K(X_1, X_2)$ denotes the matrix obtained by evaluating the covariance function $k(\mathbf{x},\mathbf{x}')$ between all pairs of points in $X_1$ and $X_2$.

The posterior covariance function is given by
\begin{equation}
k_p(x,x') = k(x,x') - K(x, X)\bigl[K(X, X) + \Sigma\bigr]^{-1} K(X, x').
\end{equation}

As for any second-order random process, the KL modes of the posterior GP can be obtained by solving the eigenvalue problem of the integral operator associated with $k_p$ (see Eq.~\ref{eq2}).

\subsection{Priors for the QCD shear viscosity}

The shear viscosity to entropy density ratio $\eta/s(T)$ is one of the key transport coefficients controlling the hydrodynamic evolution of the medium formed in HIC. It determines how efficiently momentum anisotropies are dissipated during the expansion and therefore has a direct impact on flow observables. Inferring its temperature dependence from experimental data is one of the central goals of Bayesian analyses in heavy-ion phenomenology. However, because $\eta/s(T)$ is a function rather than a finite set of parameters, its prior specification requires assumptions about smoothness, correlations, and physical constraints. We begin with a minimal GP prior for $\eta/s(T)$ and then discuss a pushforward construction that enforces positivity by construction.

\subsubsection{Minimal model}

The simplest thing one can do is draw samples for $\eta / s (T)$ directly from a GP with specified mean $m(T)$ and covariance $k(T,T')$ functions. In this way, the information contained in the prior is completely specified by the choice of these functions. A common choice for the covariance function is the Radial Basis Function (RBF)
\begin{equation}
    k(x,x') = \sigma^2 \exp\left( -\frac{|x-x'|^2}{2l^2}  \right)
\end{equation}
with hyperparameters $\sigma$ and $l$ that control, respectively, the amplitude of fluctuations and the characteristic range of correlations. Samples of a GP with zero mean and RBF covariance are infinitely differentiable, which is good for numerical calculations but can be a problem if one wants to infer functions that describe phase transitions. Other classes of covariance functions can generate paths that are not as smooth. The Matérn class of covariance functions, for example, has a parameter that controls directly how many times the samples are differentiable and can be more suited if a phase transition is known or expected. A short guide on how to choose kernels can be found in \cite{Duvenaud2014KernelCookbook}.

We start our analysis with a minimalistic GP prior for $\eta/s(T)$. We model our prior distribution as
\begin{equation}
    \eta/s(T) \sim \mathcal{GP}(m(T), k(T,T'))
\end{equation}
with constant mean function $m(T) = m$ and RBF covariance with constant hyperparameters $\sigma^2 $ and $l^2 $. The choices of $m, l$ and $\sigma$ must reflect prior knowledge about $\eta/s(T)$. In this illustrative prior, we take $m=1/(4\pi)$, although for realistic inference it could be appropriate to choose a higher value, since previous analyses indicate a considerably larger shear viscosity. The parameter $l$ determines how fast the samples vary as functions of $T$ and the value of $l = 0.2\,\mathrm{GeV}$ is chosen by hand for this work. 

The value of $\sigma^2$ corresponds to the variance of the process and determines how broad our prior is.
The choice of $\sigma^2$ is more complicated because, by construction, the variance of our process is symmetric with respect to the mean, while we need $\eta/s(T) \geq 0$ for physical consistency. In principle, one could choose a value for $\sigma$ such that the number of unphysical samples is negligible, but that could also make the prior unreasonably narrow. As a first possibility, we impose positivity ``on the fly'' by rejecting and discarding unphysical samples. Other constructions that enforce positivity more naturally, and avoid some of the drawbacks of rejection, will be discussed below. This process ultimately changes the prior distribution and makes it non-Gaussian. We choose $\sigma = 0.04$, which gives a reasonably wide prior while keeping the fraction of unphysical samples small (roughly 8\%). Positivity was evaluated on a uniform grid of 1000 points over $T\in [0,0.5]$ GeV and 7746 of $10^5$ samples were rejected.

\begin{figure}[t]
  \centering
  \includegraphics[width=\columnwidth]{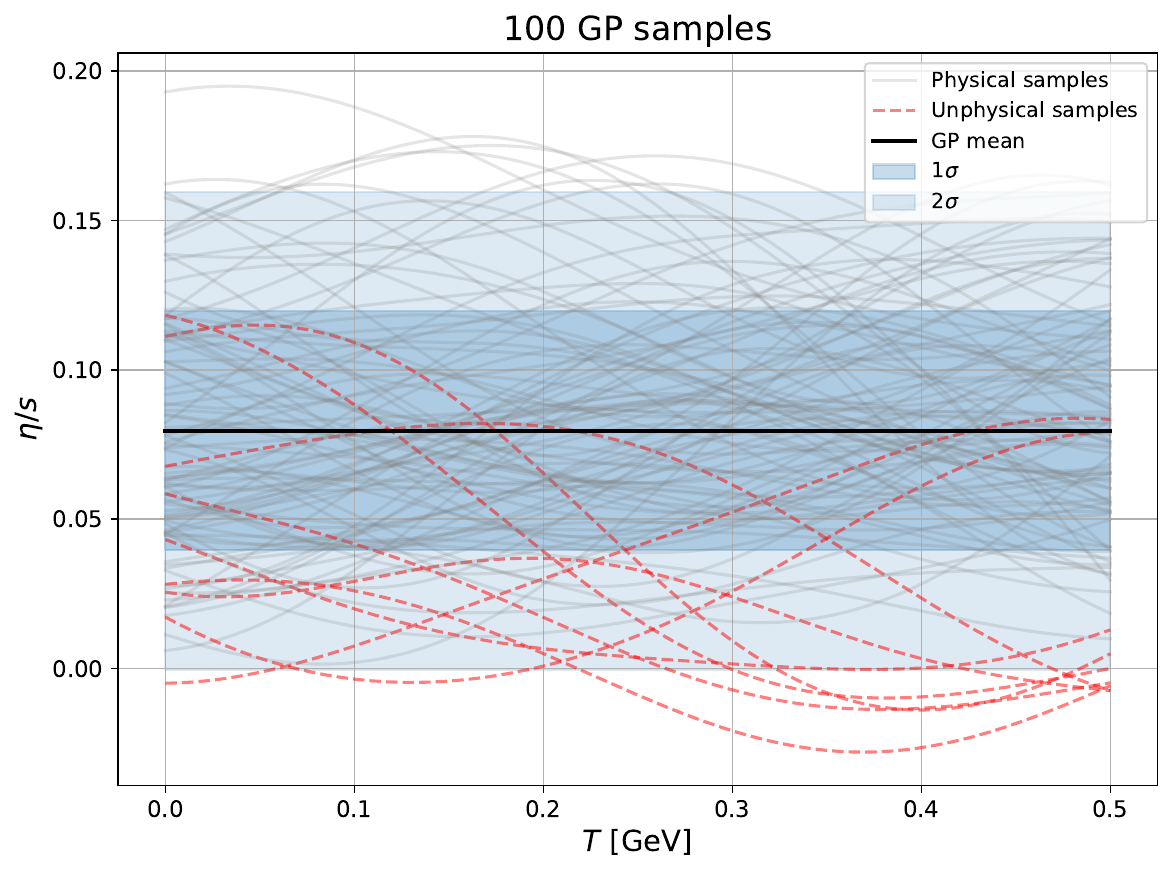}
  \caption{100 samples of the unconstrained GP prior for $\eta/s(T)$. Unphysical samples are shown in red dashed lines, and credible intervals are shown with blue bands.}
  \label{fig:minimalPrior}
\end{figure}

\begin{figure}[t]
  \centering
  \includegraphics[width=\columnwidth]{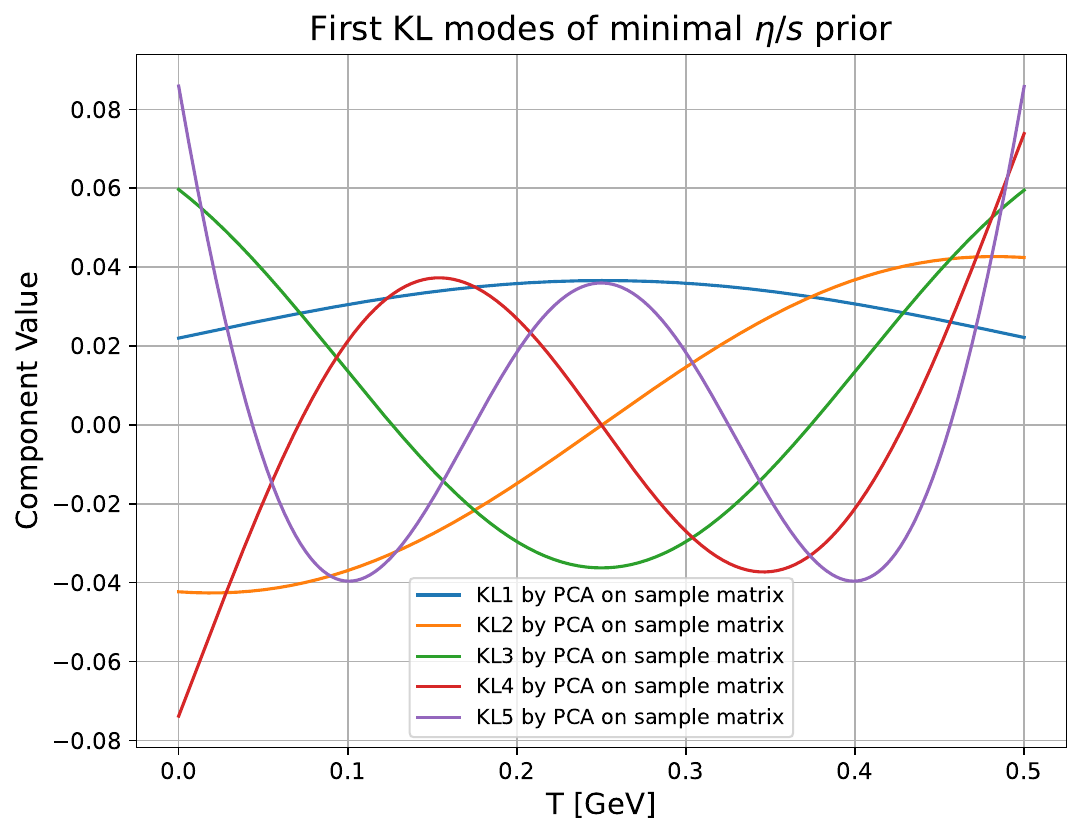}
  \caption{The first 5 KL modes of the unconstrained GP prior for $\eta/s(T)$, ordered by the magnitude of the respective eigenvalues. }
  \label{fig:minimalKLmodes}
\end{figure}

\begin{figure}[t]
  \centering
  \includegraphics[width=\columnwidth]{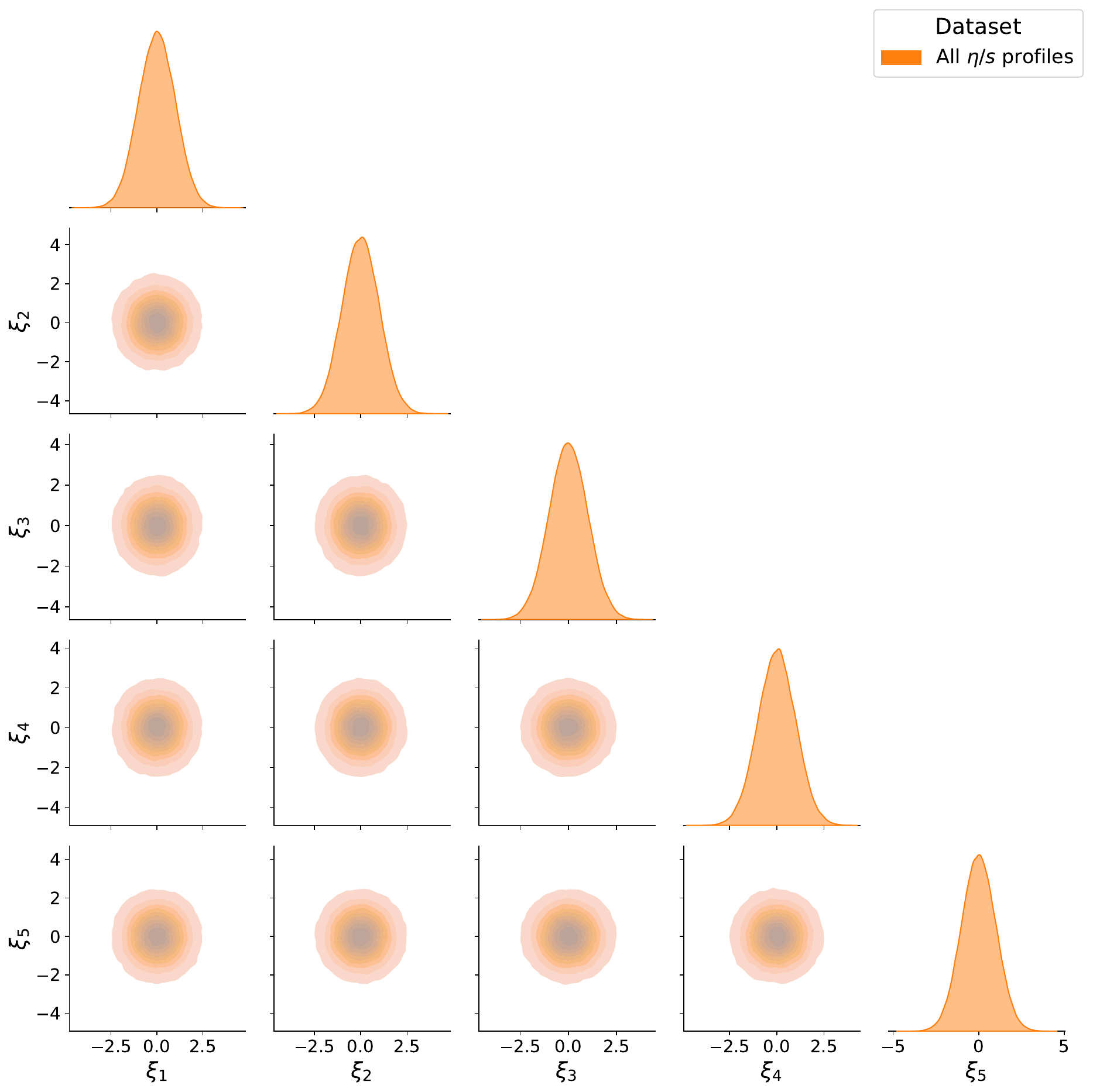}
  \caption{Corner plot of the distribution of KL coefficients for the unconstrained GP prior for $\eta/s(T)$. This plot includes physical and unphysical samples. The distributions are consistent with the expectation for a GP: i.i.d\ Gaussian variables with zero mean and unit variance. }
  \label{fig:minimalFullCorner}
\end{figure}

\begin{figure}[t]
  \centering
  \includegraphics[width=\columnwidth]{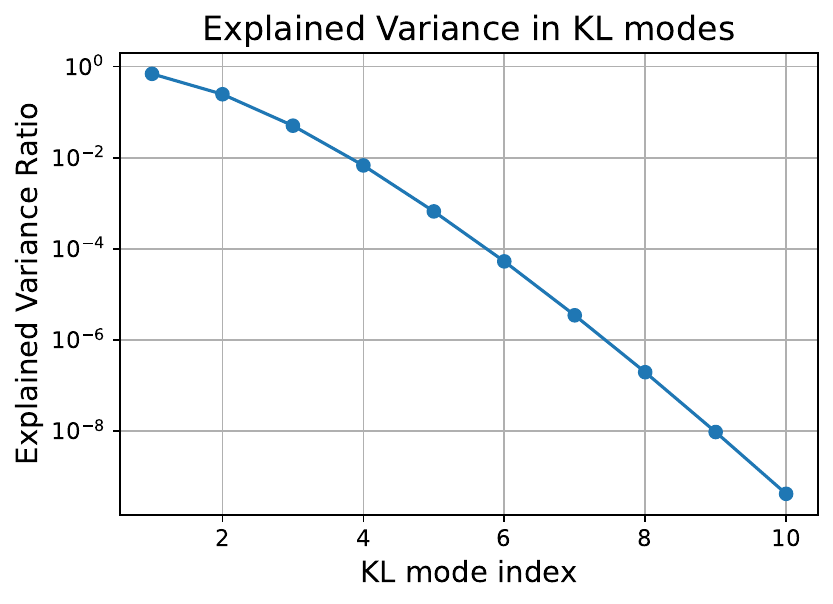}
  \caption{The normalized eigenvalues of the covariance operator for the unconstrained GP prior for $\eta/s(T)$. The spectrum shows approximately exponential decay, which allows for the construction of satisfactory finite-dimensional representations. }
  \label{fig:minimalEigenvalues}
\end{figure}

\begin{figure}[t]
  \centering
  \includegraphics[width=\columnwidth]{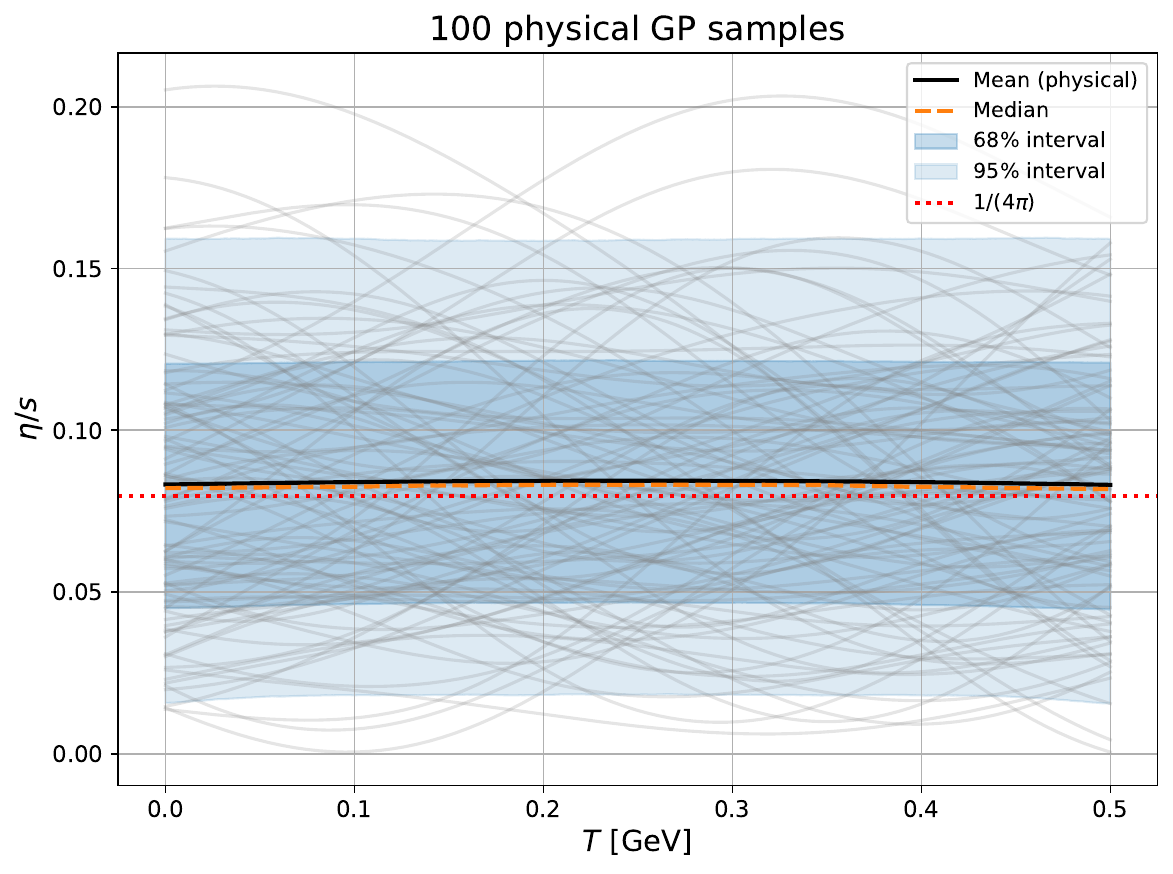}
  \caption{100 samples of the physical ensemble obtained by rejecting unphysical samples. Credible bands are shown in blue along with horizontal lines showing the mean, median and the reference value $1/(4\pi)$.}
  \label{fig:physicalMinimalPrior}
\end{figure}

\begin{figure}[t]
  \centering
  \includegraphics[width=\columnwidth]{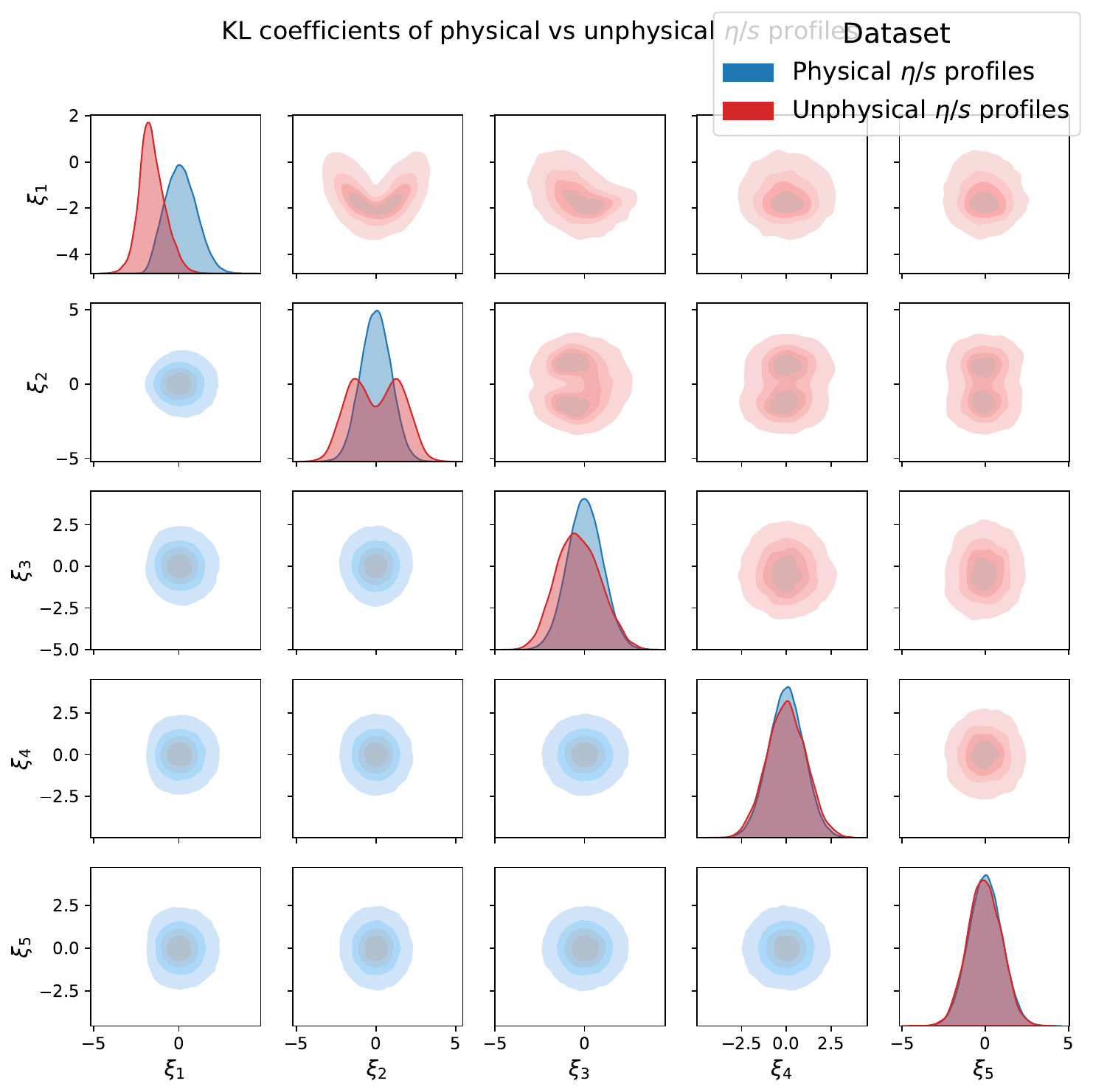}
  \caption{Corner plot of the KL coefficients for physical (blue) and unphysical (red) samples in the KL basis of the full process.}
\label{fig:minimalUnphysicalcorner}
\end{figure}

Fig. \ref{fig:minimalPrior} shows the uncertainty bands of the GP, along with 100 samples that are colored based on whether they are physical or not. The first KL modes of the minimal prior for $\eta/s$ (including unphysical samples) are shown in Fig. \ref{fig:minimalKLmodes}. The distribution of the KL coefficients for these modes is shown in the corner plot in Fig. \ref{fig:minimalFullCorner}, and the covariance eigenvalues $\lambda_i$ are shown in Fig. \ref{fig:minimalEigenvalues}. The observed distribution of coefficients is consistent with the expectation for a GP of i.i.d\ Gaussian variables $\mathcal{N}(0,1)$.

As mentioned before, discarding samples induces changes in the statistical properties of the resulting random process. It certainly breaks the Gaussianity of the process and possibly introduces correlations and higher-order statistical dependencies in the distributions of coefficients. Fig. \ref{fig:physicalMinimalPrior} shows credible bands and other statistical quantities for the physical samples, and one can clearly see, as expected, that the distribution is skewed towards the positive direction. In Fig. \ref{fig:minimalUnphysicalcorner}, the distributions of coefficients for physical and unphysical samples are shown in a corner plot (in the KL basis of the full GP).

It should also be noted that the optimality of the truncated parametrization is guaranteed only with respect to the measure of the original process from which the KL modes were computed. Since the selection of physical samples changes the covariance function of the process, to obtain a new optimal parametrization, one should compute the KL modes of the new induced covariance function, which is not known analytically. In this case, one can approximate the KL modes by using a set of samples from the process, as discussed in Appendix A. The coefficients obtained by projecting onto the recomputed KL basis are uncorrelated by construction, but they are not necessarily independent because the constrained process is non-Gaussian. In practice, we find that the KL modes and eigenvalues are essentially unchanged by the rejection of unphysical samples. Thus, the original KL expansion of the GP still provides a good parametrization of the physical ensemble in the $L^2$ sense. It also means that the covariance function of the resulting process is similar to the original one because of Eq. \ref{eq5}.

A practical issue for inference arises if one interprets the accepted physical samples as defining a new prior distribution over KL coefficients. After rejecting unphysical functions, the distribution of coefficients is no longer the original product Gaussian distribution and is not known analytically. Therefore, if one wants to sample directly from this induced physical ensemble as an explicit prior, one would need to approximate its density from samples, for example using Kernel Density Estimation or Normalizing Flows \cite{Yamauchi:2023xrz, Roch:2025jpu}.

However, this density-estimation step is not the only solution. One can instead keep the original Gaussian prior over the KL coefficients and impose physicality at the level of the likelihood evaluation during MCMC. In this approach, the prior remains the analytically known Gaussian prior associated with the unconstrained GP, while proposals that generate unphysical functions are assigned zero likelihood and rejected before evaluating the emulator. 
This formulation requires checking the physicality constraints at each MCMC proposal, but it does not require evaluating or normalizing the probability density of the rejected physical ensemble. These two formulations yield the same posterior distribution over physical functions: conditioning the Gaussian prior on the physicality constraint differs from multiplying the likelihood by the corresponding indicator function only by a normalization constant. The latter formulation retains the original Gaussian measure as a convenient reference for constructing proposals.

\subsubsection{Pushforward model}

Instead of hard rejecting unphysical samples, another reasonable approach is to implement the physical constraints directly in the construction of the random process prior. This is not a trivial task and might not be possible in every case of interest. A very powerful and widely adopted approach is to put a GP prior over an auxiliary variable and map that variable into the quantity of interest through some pushforward function. This induces a distribution over the quantity of interest which we use as a prior. The great advantage of using this pushforward construction is that our prior is exactly Gaussian over the latent auxiliary variable and we can do inference on that space, where the prior is analytically known. Moreover, Gaussian priors are particularly convenient for Bayesian inference, as some MCMC methods are explicitly constructed for Gaussian reference measures and are not directly applicable to more general priors.

\begin{figure}[t]
  \centering
  \includegraphics[width=\columnwidth]{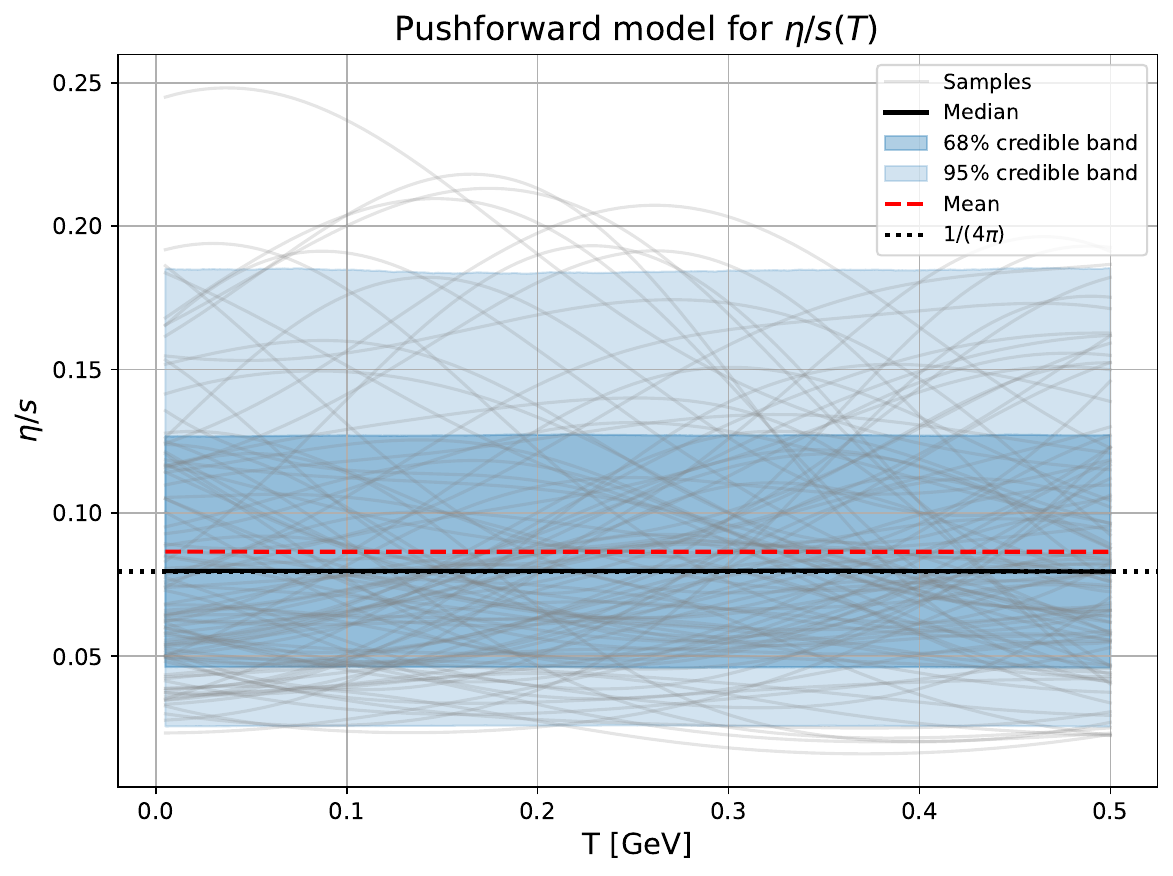}
  \caption{Pushforward prior for $\eta/s(T)$. The plot shows 100 samples along with credible bands and horizontal lines indicating the mean and median. Every sample is now guaranteed to be physical, but the distribution is skewed as can be seen by comparing the mean and median of the process at each point. The median corresponds to the reference value $1/(4\pi)$. }
  \label{fig:pushforwardPrior}
\end{figure}

To enforce positivity, we need to put our GP samples through some non-linear function $f:\mathbb{R} \rightarrow [0,\infty)$. A common choice is $f(x) = \exp x$. Another interesting choice is the softplus function $f(x) = \ln(1+\exp x)$, which we use here for the $\eta/s(T)$ prior. As discussed, we define a function $\phi(T)$ such that
\begin{equation}
    \eta/s(T) = C \ln(1+\exp{\phi(T)}) \label{softplus}
\end{equation}
and put a GP prior over it
\begin{equation}
    \phi(T) \sim \mathcal{GP}(0, k(T,T'))
\end{equation}
with 
\begin{equation}
    C = \frac{1}{4\pi \ln 2}
\end{equation}
so that the median of the $\eta/s$ prior matches the reference value $1/(4\pi)$ and $k(T,T')$ is a RBF covariance with $\sigma = \sqrt{0.5}$ and $l = 0.2\,\mathrm{GeV}$. The credible bands, mean and median are shown along with 100 samples in Fig. \ref{fig:pushforwardPrior}.

\begin{figure}[t]
  \centering
  \includegraphics[width=\columnwidth]{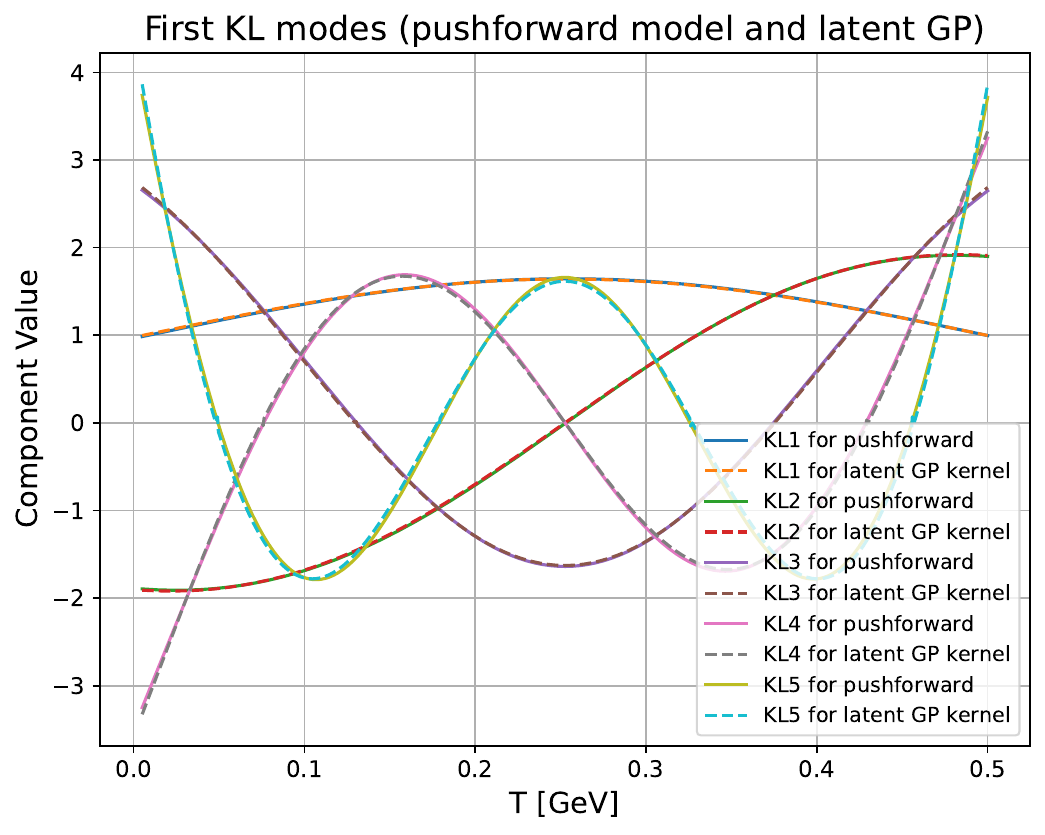}
  \caption{First KL modes of the induced pushforward process (lines) and of the latent GP (dashed). The modes are seen to be similar, indicating that the covariance structure of the latent process is mostly preserved in the pushforward process. More significant changes are observed in the distribution of KL coefficients.}
  \label{fig:pushforwardKLModes}
\end{figure}

\begin{figure}[t]
  \centering
  \includegraphics[width=\columnwidth]{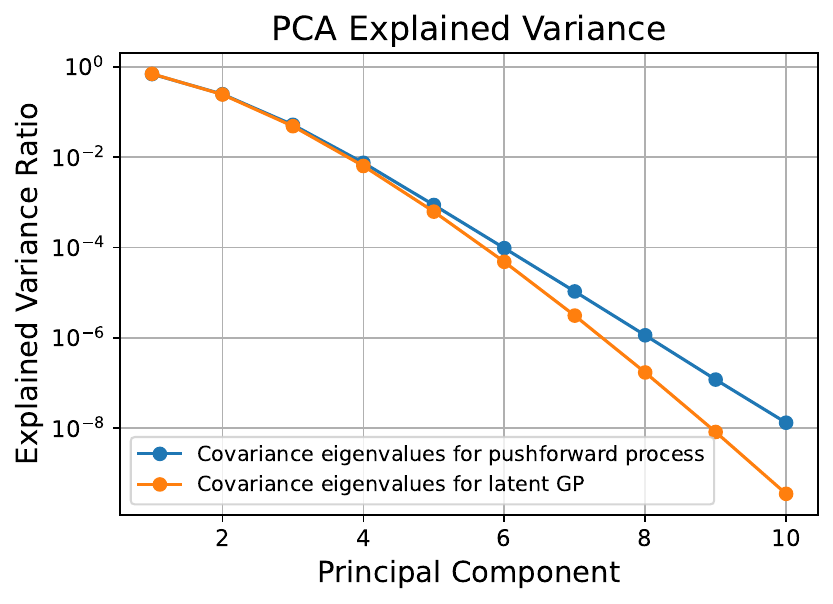}
  \caption{First normalized eigenvalues of the covariance operator for the pushforward and latent processes. Both show approximately exponential decay and the variation in the spectra between the two processes is observed to be small.}
  \label{fig:pushforwardSpectra}
\end{figure}

\begin{figure}[t]
  \centering
  \includegraphics[width=\columnwidth]{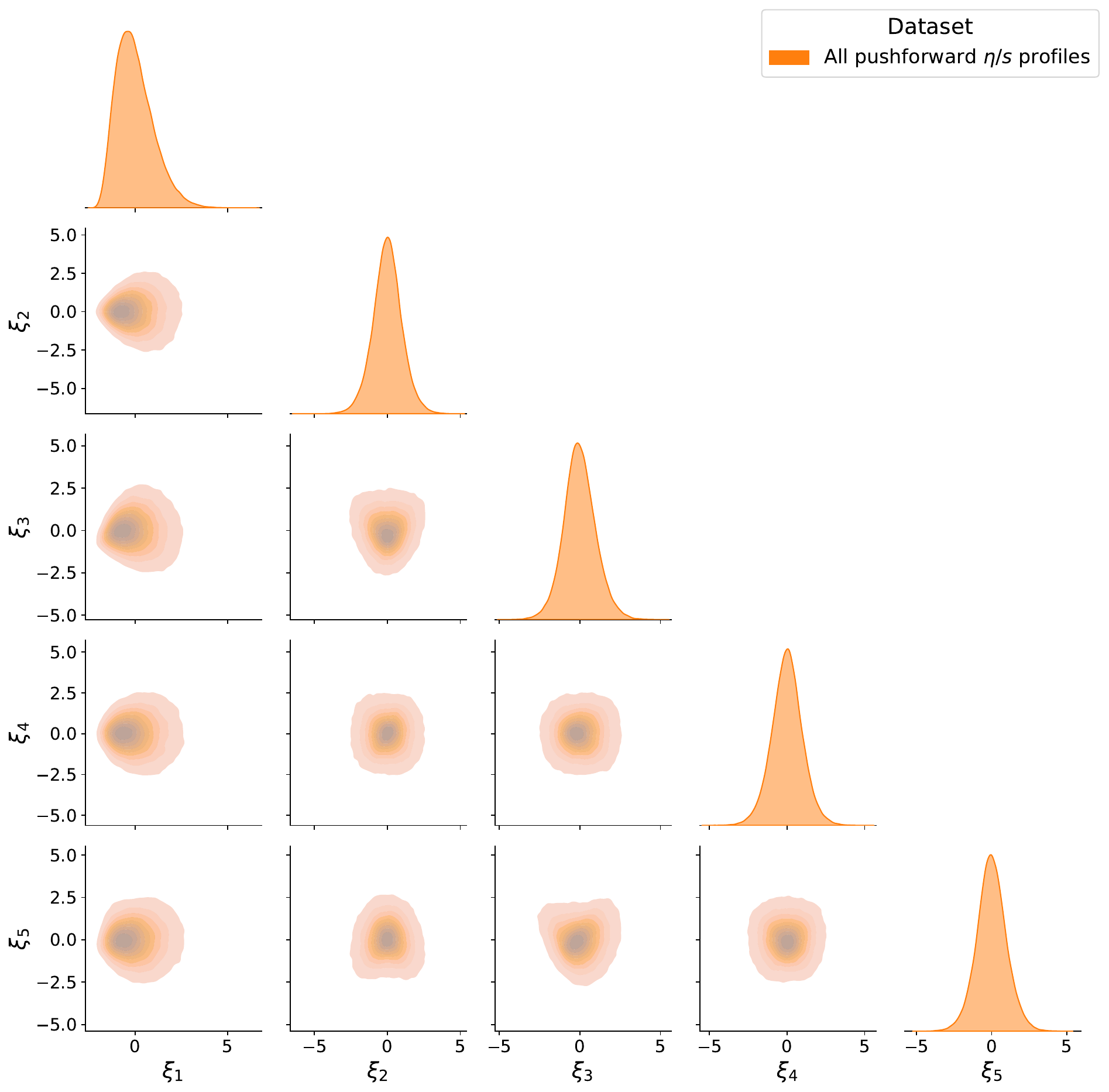}
  \caption{Corner plot for the distribution of first 5 KL coefficients of the pushforward process for $\eta/s(T)$. All samples are physical by construction, but the coefficients are not independent as in the case of a GP.}
  \label{fig:pushforwardCorner}
\end{figure}

The pushforward process also admits a KL expansion, obtained by diagonalizing the induced covariance function, but since it is non-Gaussian the KL coefficients are not independently distributed. The KL modes of the induced and latent processes are shown in Fig. \ref{fig:pushforwardKLModes}, and spectra of the latent and induced process are shown in Fig. \ref{fig:pushforwardSpectra}. The KL modes of the induced process for $\eta/s(T)$ are similar to the modes of $\phi$, and so is the spectrum. Clearer although still small changes can be seen in the distribution of coefficients in the corner plot in Fig. \ref{fig:pushforwardCorner}, demonstrating that the modes are no longer independent. The fact that the covariance structure of the process is not severely altered by the pushforward can be understood as a consequence of the fact that the pushforward function is close to linear in the typical range of values for $\phi$, so the non-linearity ends up affecting only the tails of the Gaussian distributions, which decay fast. This is also a reason why the pushforward in Eq. \ref{softplus} is arguably better than others, such as the exponential map $\eta/s(T) = \exp{\phi(T)}$, which have the correct domain and codomain but are more non-linear in the typical range of values of $\phi$.

If one were to do inference over the KL coefficients of the pushforward process, the prior distribution must account for those dependencies. As said before, the advantage of the pushforward prior is that the latent function $\phi$ is distributed as a GP and we can do inference directly over $\phi$ because the prior distribution of KL coefficients is simple there. We stress, however, that the KL basis is $L^2$ optimal for the representation of samples of the associated process, and putting the samples through a pushforward affects that optimality. We thus expect to minimize the average error squared on $\phi(T)$, but not on $\eta/s(T)$. In practice, optimizing the representation of $\phi$ will result in representations of $\eta/s(T)$ that are close to optimal, but a truly optimal representation for the pushforward process comes with the price of having a complicated prior distribution inference. In principle, a representation obtained by truncating the KL expansion of the latent process and then applying the pushforward map can outperform the “optimal” representation given by the KL expansion of the resulting process. This is because the KL optimality is defined within the class of linear parametrizations, whereas the pushforward induces a non-linear parametrization.

\subsection{Priors for the QCD EoS}
The Equation of State (EoS) is one of the crucial functions for describing the evolution of the medium formed in HIC. There is great interest in methods for inferring the EoS from data
because the sign problem severely limits first-principles lattice-QCD
calculations at finite baryon chemical potential $\mu_B$. Heavy-ion collisions at top LHC energies are very close to $\mu_B = 0$, but collisions at intermediate energies (such as in the Beam Energy Scan program) and other systems (most notably neutron stars) are not. A deeper understanding of the QCD phase diagram necessarily involves knowing the EoS in these more general situations. We analyze the EoS prior proposed in \cite{gong2024gaussianprocessgenerativemodel} and then briefly discuss how fully physical EoS priors could alternatively be constructed using pure pushforward methods.

The use of flexible, probabilistic representations of equations of state has a long history in neutron-star applications. In particular, nonparametric Gaussian-process constructions have been used to infer the cold dense-matter equation of state from gravitational-wave and electromagnetic observations \cite{Landry:2018prl,Essick:2019ldf,Landry:2020vaw}, while related Bayesian Gaussian-process models have been developed to quantify correlated chiral-EFT truncation uncertainties and propagate them to thermodynamic quantities such as the pressure and speed of sound \cite{Drischler:2020hwi}. More recently, Gaussian-process model mixing has been used to combine low-density chiral-EFT information with high-density perturbative-QCD constraints in symmetric nuclear matter \cite{Semposki:2024vnp}. These works are conceptually relevant to the present study in that they treat the equation of state as a random function and propagate this uncertainty through thermodynamic relations. Here, however, we focus on the finite-temperature QCD equation of state relevant to heavy-ion collisions. We first analyze the construction proposed by Gong et al. and then discuss how a pushforward from latent random functions with a Gaussian prior could provide a fully physical alternative.

For $\mu_B = 0$, the EoS can be represented by a one-dimensional function $P(T)$ that relates the pressure $P$ and local temperature $T$. Often, one defines the scaled pressure $\tilde{P}(T) = P(T)/T^4$ to make it dimensionless. 

The construction of the prior distribution proposed in \cite{gong2024gaussianprocessgenerativemodel} starts by regressing a GP over two sets of points: a low-temperature set ($T < 0.13\,\mathrm{GeV}$) obtained from the Hadron Resonance Gas (HRG) model and a high-temperature set ($T\geq0.7\,\mathrm{GeV}$) from lattice calculations \cite{Bazavov_2014}. The GP is built with a RBF kernel with fixed $\sigma^2 = 1$ and optimized $l = 0.473$.
To have physical samples, one must impose 4 constraints in the random process that generates the EoS samples:
\begin{equation}
    P(T) > 0,\quad
    \frac{\partial P}{\partial T} > 0,\quad
    \frac{\partial^2 P}{\partial T^2} > 0 \;\quad
     \text{ and }\quad 0 \leq c_s^2 < 1 . 
     \label{eq11}
\end{equation}
The first 3 are related to thermodynamic consistency, while the last one enforces causality. Following \cite{gong2024gaussianprocessgenerativemodel}, we additionally impose $c_s^2<1/2$, a stricter bound than the fundamental causal requirement $c_s^2<1$.
The speed of sound squared $c_s^2$ is obtained from a particular sample of $P(T)$ as
\begin{equation}
    c_s^2 = \frac{\partial P}{\partial e} = \frac{\partial P / \partial T}{\partial e / \partial T} = \frac{\partial P / \partial T}{T \, \partial^2 P / \partial T^2}.
\end{equation}
The first constraint in \ref{eq11} is imposed on the construction of the random process by defining the GP samples as relations between $\ln(\tilde{P})$ and $\ln(T/T_0)$, so that a sample of $\ln(\tilde{P})$ as a function of $\ln(T/T_0)$ uniquely determines a sample of $P(T)$ that is ensured to be positive everywhere (this is similar to enforcing positivity with a pushforward as done for the shear viscosity, but with an extra transformation to the GP argument). The authors use $T_0 = 1\,\mathrm{GeV}$. The last three conditions are imposed ``on the fly'', that is, by sampling and rejecting samples that don't satisfy one or more of the constraints. 

There are two important things to notice. First, defining the GP in log space induces a random process over $\tilde{P}(T)$ that is not Gaussian. It can be classified, by definition, as a lognormal random process. Secondly, as seen before, imposing constraints on the fly changes the probability distribution of the final random process. This will, in general, break the Gaussianity of the process over $\ln(\tilde{P})$ and consequently the ``lognormality'' of the induced process over $\tilde{P}(T)$. In the context of using random processes as prior distributions for Bayesian analysis, it is essential to understand, in a quantitative manner, how the probability distribution over function space changes as physical constraints are imposed. This will ensure that the knowledge we want to incorporate into the prior distribution is correctly accounted for when the inference step is performed. 

\begin{figure}[t]
  \centering
  \includegraphics[width=\columnwidth]{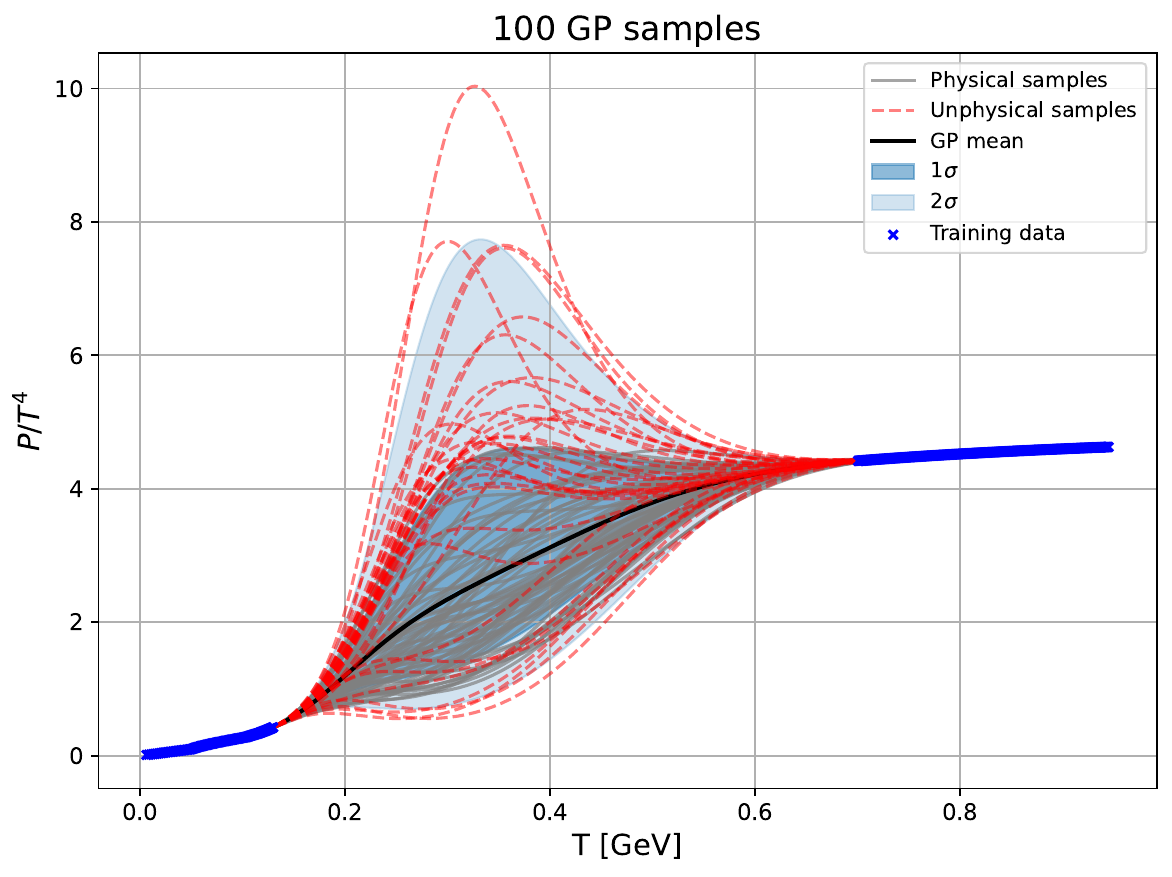}
  \caption{100 samples of Gong et al. prior for the EoS. Unphysical samples are shown in red, and the credible intervals are shown in blue. The credible intervals refer to the full process, before rejecting unphysical samples.}
  \label{fig:GongPrior}
\end{figure}

Fig. \ref{fig:GongPrior} shows 100 samples from the prior constructed by Gong et al. Around $30\%$ of the resulting samples are unphysical because they violate one or more of the last three constraints in \ref{eq11}. The constraints were evaluated using second-order finite differences on 1000 points uniformly spaced in $\ln(T/T_0)$ and 21,538 of 71,538 tested samples were rejected. State-of-the-art numerical simulations may fail or produce thermodynamically ill-posed evolutions when fed an EoS that violates these conditions, so these samples cannot be consistently used to evaluate a likelihood. Besides that, as argued in the construction of shear viscosity prior, imposing physical constraints by rejection sampling effectively breaks the Gaussianity of the resulting prior, even in the latent variable where the GP is defined. Pure pushforward constructions are advantageous in this aspect because they retain a latent random function with an exactly Gaussian measure, where inference can be performed with enhanced tractability. 

If one nevertheless wishes to use rejection-based constructions while retaining the Gaussian measure as the reference measure of the sampling formulation, a pragmatic approach is to impose the physical constraint at the likelihood level rather than in the prior itself. Emulators can be trained on a set of strictly physical samples, and at each step of the Markov chain, before evaluating the likelihood, one checks whether the current proposal is physical. If the proposal is unphysical, it is assigned zero likelihood and immediately rejected.
This avoids evaluating the emulator in regions of parameter space where the model is not well defined. Since simulations are expected to fail there, emulator predictions would amount to unphysical extrapolations of the underlying model. Provided that the physicality test is sufficiently cheap and does not make inference impractically slow, this strategy allows hard constraints to be enforced via rejection while preserving some of the tractability advantages of Gaussian priors.

\begin{figure}[t]
  \centering
  \includegraphics[width=\columnwidth]{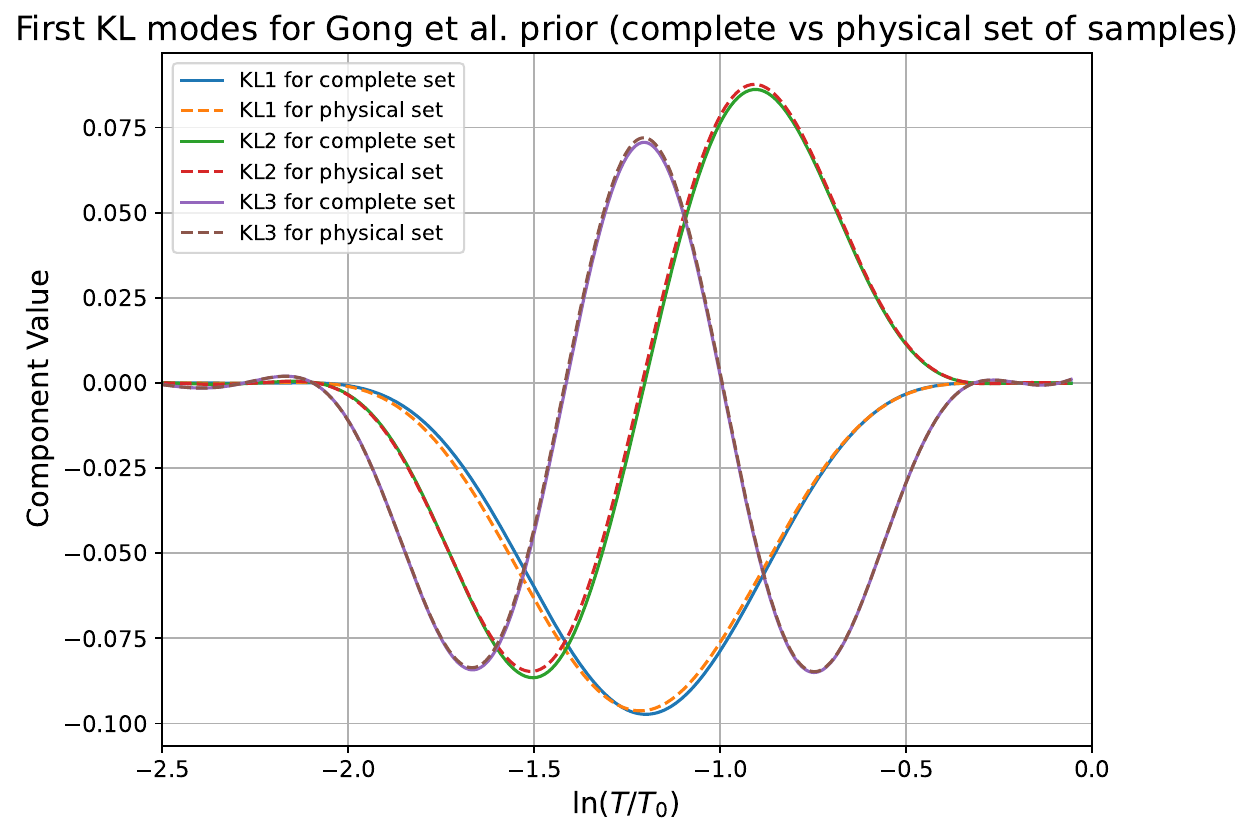}
  \caption{The continuous lines show the first KL modes of the Gaussian process used by Gong et al. to construct a prior for the EoS. Most notably, the modes are close to zero in the regions where the GP was trained on data. Dashed lines show the KL modes of the (non-Gaussian) prior obtained by rejecting unphysical samples.}
  \label{fig:GongKLmodes}
\end{figure}

\begin{figure}[t]
  \centering
  \includegraphics[width=\columnwidth]{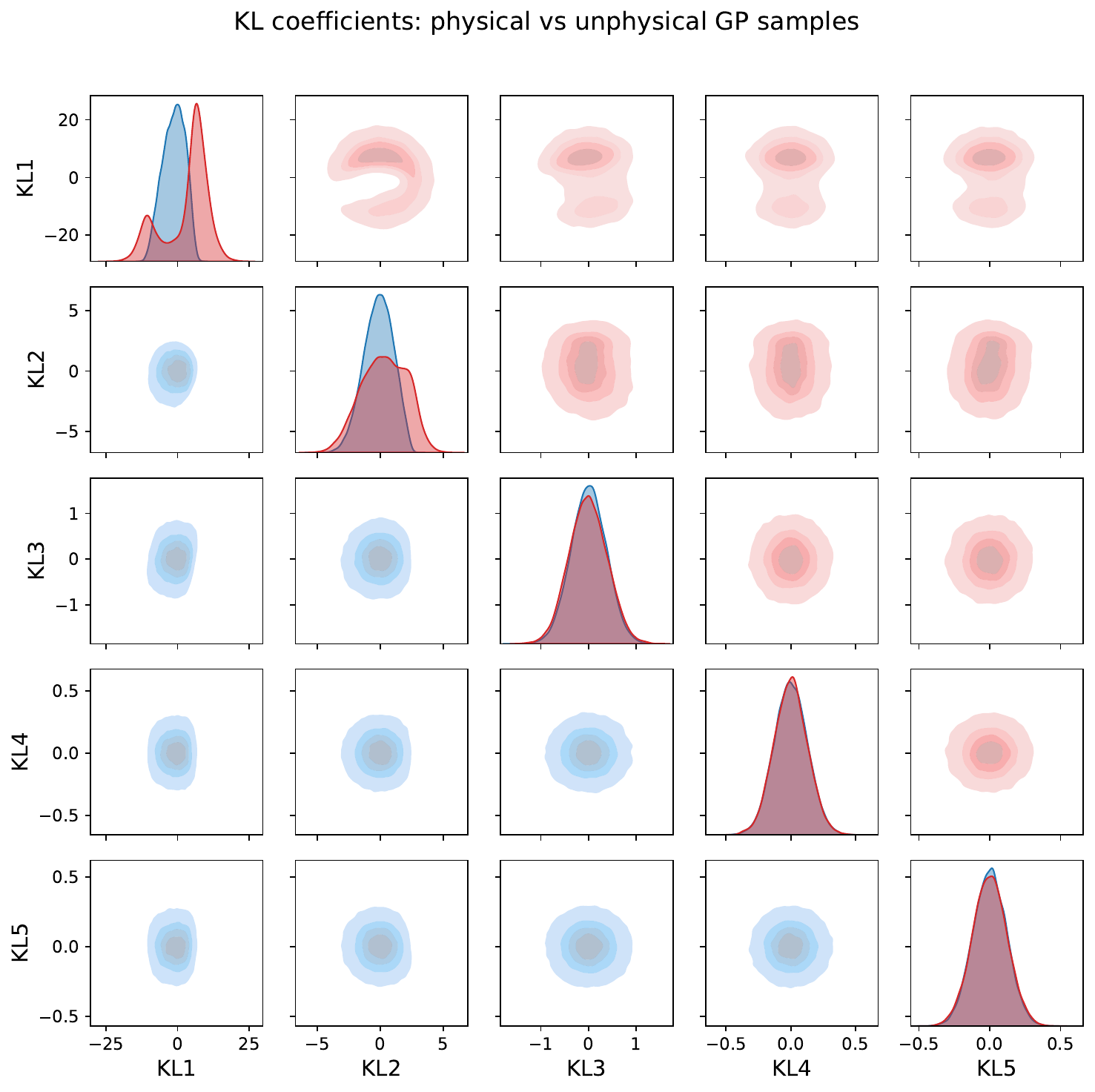}
  \caption{Corner plot of the KL coefficients of the physical (blue) and unphysical (red) samples in the KL basis of the full GP used by Gong et al.}
  \label{fig:unphysicalCornerGongEoS}
\end{figure}

Fig. \ref{fig:GongKLmodes} shows the first 3 KL modes for the latent GP in the EoS construction and the first 3 KL modes of the physical prior obtained by rejecting unphysical samples. The modes look like wavelets of increasing frequency and are very close to zero in the regions where the GP is constrained (high and low temperature regions). The rejection of unphysical samples is seen to have little effect on the KL modes: its effect on the covariance of the process is mostly expressed through changes in the distribution of KL coefficients rather than on the modes. Fig. \ref{fig:unphysicalCornerGongEoS} shows a corner plot for the distribution of the first 5 KL coefficients (in the full process basis) for physical and unphysical sets. The distribution of coefficients in the full GP is by construction i.i.d\ Gaussian variables of unit variance. This is not true for the physical set of samples.

It is also interesting to note that when converting the samples to functions of $T$ instead of its logarithm the modes get distorted in a way that compresses low temperature intervals more than high temperature ones, resulting in modes that oscillate slower as the temperature increases (in contrast to what Fig. \ref{fig:GongKLmodes} shows, where each mode oscillates with a constant period). We also verified, as in earlier examples, that the variance eigenvalues decay approximately exponentially, supporting the use of low-dimensional parametrizations.

Finally, although we do not construct such a prior explicitly here,
fully physical EoS priors can also be constructed using pure pushforward
methods. Instead of sampling the pressure or interaction measure directly and rejecting unphysical samples, one may sample an unconstrained latent function and map it to a manifestly physical quantity, such as the speed of sound $c_s^2(T)$. By choosing a transformation that enforces stability and causality, and then reconstructing the thermodynamic functions through the $\mu=0$ thermodynamic identities, the resulting pressure, entropy density and energy density automatically satisfy the desired consistency conditions.

Constructions of this type have been developed and used in the neutron-star equation-of-state literature \cite{Landry:2018prl, Landry:2020vaw, Miller:2021qha}. They illustrate that pushforward priors can be designed to produce ensembles that are physical by construction, with no need for rejection sampling. In the context of heavy-ion phenomenology, this provides a possible route toward EoS priors with controlled smoothness and correlation structure, while maintaining a fully physical prior ensemble.

\subsection{Implications for emulator-based inference with realistic simulations}

We do not perform a full calibration study in this work. In the present section, we discuss how the prior constructions introduced above interface with the standard emulator-based inference workflow used in heavy-ion phenomenology, and we highlight the practical implications of nonparametric functional priors for surrogate modeling and MCMC sampling. 

The standard Bayesian calibration problem in HIC phenomenology is to infer model parameters $x$ from experimental observables $y$, related by a computationally expensive forward model $G$ as $y = G(x)$. Because no closed-form relationship exists between $x$ and $y$, the forward problem is solved numerically, and evaluating $G$ at a single point in parameter space is already too costly to be done at every step of an MCMC random walk. The HIC community addresses this by training an emulator, typically a GP regressed over a Latin hypercube design in parameter space, that interpolates simulation outputs across the full parameter space at negligible computational cost. Likelihood evaluations during MCMC then query the emulator rather than the underlying simulation. State-of-the-art global analyses typically operate in parameter spaces of dimension 5 to 20, and common parametrizations of $\eta/s(T)$ use up to 4 parameters. The priors adopted in these analyses are almost exclusively uniform distributions, with the exception of \cite{Heffernan_2024}, which employs generalized Normal distributions.

The functional priors constructed in this work can be incorporated into this pipeline in two ways depending on the parametrization used. The pushforward parametrization for $\eta/s$ is the most straightforward to integrate, since the $N=3$ truncation captures over $99\%$ of the prior variance, $N=4$ over $99.9\%$, and the marginal distribution of each KL coefficient is known in closed form and can be passed directly to any standard MCMC sampler. For the parametrizations constructed with on-the-fly rejection sampling, the prior over coefficients is not available analytically, and additional work is needed to make it usable for MCMC. One natural approach is to train a Normalizing Flow on samples from the rejection procedure, yielding a tractable density that approximates the true prior. Uncertainty from truncated modes can be incorporated by marginalizing over their prior or by retaining additional modes in the inference.

\subsubsection{Function-space MCMC algorithms}

Functions are naturally infinite-dimensional objects, so ideally inference should be performed directly in infinite-dimensional space. In practice, some discretization is always necessary, which restricts the originally infinite-dimensional prior to a finite-dimensional subspace on which the inference is ultimately carried out. The natural goal is then to choose subspaces that capture as much variance as possible from the full infinite-dimensional prior, and this is precisely what truncated KL expansions are designed to achieve.

Most standard MCMC algorithms, such as Random Walk Metropolis, suffer from the curse of dimensionality: their efficiency degrades as the dimension of the parameter space increases, requiring proposal step sizes to shrink in order to maintain reasonable acceptance rates \cite{10.1214/aoap/1034625254}. This leads to increasingly slow exploration of the posterior as the number of parameters grows. While state-of-the-art HIC analyses typically operate in parameter spaces of dimension 5 to 20, the number of KL modes required to capture a fixed fraction of the prior variance for more complex functions can be large enough to make standard MCMC methods impractical. To solve this issue, mathematicians developed dimension-robust MCMC methods, which maintain stable acceptance rates under increases in the dimension of the inference subspace. One such algorithm is the preconditioned Crank--Nicolson (pCN) method \cite{Cotter_2013}, which is specifically designed for Bayesian inference with Gaussian prior measures and is therefore directly applicable to inference over the latent GP in the pushforward parametrization of $\eta/s(T)$ constructed in this work. The same principle could, in principle, be extended to GP-based equation-of-state constructions used in neutron-star studies \cite{Landry:2018prl,Landry:2020vaw,Miller:2021qha}, provided that their priors are represented explicitly in terms of the latent Gaussian process rather than only through the generated ensemble of physical EoS samples.
 With pCN, one could include a larger number of KL modes in the inference even when the data is expected to constrain only the lowest modes, since the algorithm does not degrade as additional modes are included. This also makes the treatment of truncation uncertainty more principled: rather than fixing $N$ and discarding the remaining modes, one can retain them in the sampler and allow the data to determine how many modes are actually informed by the likelihood. The pCN method has the further conceptual advantage that the random walk is defined in function space, with the finite-dimensional parametrization entering only through the likelihood evaluation at each acceptance step.

Related preconditioned sampling strategies have recently appeared in heavy-ion Bayesian analyses. For example, \cite{Jahan:2024wpj,Jahan:2025cbp} use the \texttt{pocoMC} implementation of Preconditioned Monte Carlo, in which a normalizing flow is used to decorrelate the parameters before sequential Monte Carlo sampling. These applications, however, still concern ordinary finite-dimensional parameter spaces of phenomenological model parameters. The situation considered here is conceptually different: pCN is not merely a way of improving sampling efficiency in a high-dimensional vector space, but a proposal mechanism defined at the level of the underlying Gaussian measure on function space. The finite-dimensional KL parametrization then appears only as a practical approximation used to evaluate the likelihood.

The adoption of dimension-robust algorithms in heavy-ion phenomenology would therefore enable a more principled approach to Bayesian inference over functions, and would open the possibility of performing inference directly on classes of models that do not admit a natural low-dimensional parametrization, such as functions specifying the initial conditions of HIC or the finite-density QCD equation of state. We leave the implementation of such algorithms within the HIC inference pipeline to future work.

\section{Conclusion}

In this work, we investigated the use of random process priors to represent functions of interest in HIC phenomenology, with a view toward their use in Bayesian inference, focusing on the temperature dependence of the shear viscosity $\eta/s(T)$ and the QCD equation of state. The central problem we address is how to encode genuine prior knowledge about physical functions in a form that is both statistically principled and compatible with realistic inference pipelines. This requires facing a tension between physical consistency and computational tractability.

Gaussian processes provide a natural and flexible framework for this task, allowing prior information to be encoded with fine control over smoothness and correlations through the choice of covariance function. The Karhunen–Loève expansion provides a principled way to construct finite-dimensional parametrizations, yielding linear representations that minimize mean squared approximation error for a fixed number of parameters. This construction defines a natural baseline for inference over functions, but the key difficulty is how physical constraints modify this picture.

We showed that imposing constraints through hard rejection of unphysical samples, while conceptually straightforward, generically induces non-Gaussian measures and complicates both the specification of the prior and its use in MCMC algorithms. In contrast, pushforward constructions with latent Gaussian processes preserve the Gaussian measure of the underlying latent functions, enabling inference with analytically tractable priors and well-defined statistical structure, at the cost of no longer guaranteeing strict optimality of the induced parametrization. This trade-off has direct consequences for inference: the Gaussian structure of the pushforward priors makes them naturally compatible with dimension-robust MCMC algorithms such as pCN, which maintain stable acceptance rates as the number of parameters increases and provide a more appropriate framework for inference over function spaces than standard methods. Understanding this trade-off quantitatively is therefore essential for designing priors that faithfully encode physical knowledge without becoming intractable in practice, and more broadly highlights the importance of analyzing how physical constraints reshape the statistical properties of random process priors.

The parametrizations constructed in this work are directly compatible with current Bayesian pipelines in HIC phenomenology. At the same time, the framework developed here opens a path toward inference over functions that do not admit natural low-dimensional parametrizations, including the finite-density QCD equation of state and functions specifying the initial conditions of heavy-ion collisions, where the number of required modes may render traditional MCMC approaches impractical. We leave the implementation of such algorithms within realistic HIC inference pipelines, and the full calibration studies they could enable, to future work.

\hspace{0.5cm}

\begin{acknowledgments}

The authors would like to thank S. Mak, H. Roch and D. Mroczek for helpful discussions. G.G. was supported by the 
Coordenação de Aperfeiçoamento de Pessoal de Nível Superior - 
Brasil (CAPES) - Finance Code 001, M.L. was supported by the 
São Paulo Research Foundation (FAPESP) under projects 
2018/24720-6, 2020/04867-2, and 2023/13749-1, and L.P. was supported by the Conselho Nacional de Desenvolvimento Científico e Tecnológico (CNPq) under contract No. 132732/2025-4.

\end{acknowledgments}

\bibliographystyle{apsrev4-2}
\bibliography{biblioKL.bib}

\appendix
\section{Numerical approximation of KL modes}

To compute the KL modes of a given process, one needs to solve the eigenproblem in Eq. \ref{eq2}. The main ingredient we need for that is a covariance function $k(s,t)$, but the knowledge one has about the covariance function might differ from case to case. In some cases, the covariance function is exactly known, but sometimes only a finite set of samples is available, from which we can estimate an empirical covariance \cite{griebel2021numericalapproximationkarhunenloeveexpansion}.

For a uniform grid with spacing $\Delta t$, we first solve the
ordinary matrix eigenvalue problem
\begin{equation}
    \sum_{j=1}^{n} K_{ij}v_j^{(k)}
    = \mu_k v_i^{(k)},
    \qquad K_{ij}=k(t_i,t_j),
    \label{eqA1}
\end{equation}
where the eigenvectors are normalized,
$\sum_i v_i^{(k)}v_i^{(l)}=\delta_{kl}$. The constant quadrature
factor $\Delta t$ does not affect the eigenvectors and is therefore
omitted from the matrix eigenproblem. The corresponding approximations
to the continuum covariance eigenvalues and $L^2$-normalized
eigenfunctions are
\begin{equation}
    \lambda_k \simeq \Delta t\,\mu_k,
    \qquad
    e_k(t_i)\simeq\frac{v_i^{(k)}}{\sqrt{\Delta t}}.
    \label{eqA2}
\end{equation}
Indeed, these modes satisfy
$\Delta t\sum_i e_k(t_i)e_l(t_i)\simeq\delta_{kl}$.

When the covariance kernel is known away from the grid, the
Nystr\"om extension becomes
\begin{equation}
    e_k(t)
    \simeq
    \frac{1}{\mu_k\sqrt{\Delta t}}
    \sum_{j=1}^{n}k(t,t_j)v_j^{(k)}.
    \label{eqA3}
\end{equation} 
The advantage of nonparametric/functional methods is not to avoid discretization, but to use it in a way that the methods and results are robust with respect to the arbitrary choice of grid size. This is often referred to as dimension-robustness or invariance under mesh-refinement in the mathematical inverse problem literature \cite{Stuart_2010, Cotter_2013,dashti2015bayesianapproachinverseproblems}.

If the functional form of $k(s,t)$ is known, then all that needs to be done is to solve the eigenvalue problem. This will produce a set of eigenvectors which represent the KL modes in the chosen grid, along with the corresponding variance eigenvalues.
If the functional form of 
$k(s,t)$ is not known, one can still numerically estimate the KL modes by performing standard PCA on the sample matrix. In this case, the covariance matrix $K_{ij}$
 is replaced by its empirical estimate constructed from the samples. This procedure is justified by the fact that PCA diagonalizes the empirical covariance matrix, while the KL decomposition corresponds to the eigen-decomposition of the covariance operator. The empirical covariance converges to the true covariance as the number of samples increases, and the discrete eigenproblem converges to the continuum one as the grid is refined. In this sense, the two approaches become equivalent in the limit of infinite samples and vanishing grid spacing.

In our tests, the two methods give consistent results if a reasonable number of samples is used in the second one. Estimating the covariance empirically is the only practical way to compute KL modes in the priors constructed based on rejection strategy as described in this work, because only the covariance of the full process is known.

\end{document}